\documentclass[conference]{IEEEtran}
\usepackage{graphicx}
\usepackage{float}
\usepackage{geometry}
\usepackage{comment}
\usepackage{csquotes}
\usepackage{makecell}
\usepackage{lineno}
\usepackage{amsmath,amssymb,amsfonts} 
\usepackage{algorithmic}
\usepackage[linesnumbered,ruled]{algorithm2e}
\usepackage{float} 
\usepackage{textcomp} 
\usepackage{eso-pic}
\usepackage{xcolor}
\usepackage{subfigure}
\usepackage[english]{babel}
\usepackage[utf8]{inputenc}
\makeatletter
\def\ps@IEEEtitlepagestyle{
  \def\@oddfoot{\mycopyrightnotice}
  \def\@evenfoot{}
}
\def\mycopyrightnotice{
  {\footnotesize xxx-x-xxxx-xxxx-x/xx/\$31.00~\copyright~2023 IEEE\hfill} 
  \gdef\mycopyrightnotice{}
}

\@ifundefined{showcaptionsetup}{}{
 \PassOptionsToPackage{caption=false}{subfig}}
\usepackage{subfig}
\makeatother

\usepackage{eso-pic}

\begin{document}

\title{SMARD: A Cost Effective Smart Agro Development Technology for Crops Disease Classification}


\author{\IEEEauthorblockN{Tanoy Debnath, Shadman Wadith, and Anichur Rahman*}

\IEEEauthorblockA{\textit{Department of Computer Science and Engineering} \\
\textit{Green University of Bangladesh}, and \\
\textit{National Institute of Textile Engineering and Research (NITER), constituent institute of University of Dhaka}\\
Email: tanoy@cse.green.edu.bd, shadman@cse.green.edu.bd, and
anis.mbstu.cse@gmail.com}}

\maketitle

\begin{abstract}
\boldmath
Agriculture has a significant role in a country's economy
 The \enquote{SMARD} project aims to strengthen the country's agricultural sector by giving farmers with the information and tools they need to solve common difficulties and increase productivity. The project provides farmers with information on crop care, seed selection, and disease management best practices, as well as access to tools for recognizing and treating crop diseases. Farmers can also contact the expert panel through text message, voice call, or video call to purchase fertilizer, seeds, and pesticides at low prices, as well as secure bank loans. The project's goal is to empower farmers and rural communities by providing them with the resources they need to increase crop yields. Additionally, the \enquote{SMARD} will not only help farmers and rural communities live better lives, but it will also have a good effect on the economy of the nation. Farmers are now able to recognize plant illnesses more quickly because of the application of machine learning techniques based on image processing categorization. Our experiments' results show that our system \enquote{SMARD} outperforms the cutting-edge web applications by attaining 97.3\% classification accuracy and 96\% F1-score in crop disease classification. Overall, our project is an important endeavor for the nation's agricultural sector because its main goal is to give farmers the information, resources, and tools they need to increase crop yields, improve economic outcomes, and improve livelihoods. 

\end{abstract}


\section{Introduction}
The agricultural sector, which is essential for ensuring Bangladesh's 160 million people have access to food, substantially influences the country's economic growth \cite{nobel2024categorization}. A sizable portion of crops are lost each year to disease alone.  However, a big issue facing the agricultural industry is the low yield of different crops as a result of crop diseases brought on by microscopic fungi, bacteria, viruses, nematodes, or worms that cannot be seen without a microscope \cite{rahman2022sdn}. Early detection of plant illnesses can thereby improve agricultural disease detection and suppression systems. In order to decrease environmental and health concerns and boost crop productivity, eco-friendly techniques of controlling crop diseases and proper pesticide use should be promoted. Prioritizing research on integrated pest control and providing farmers with access to the newest technologies are essential for achieving this \cite{rahman2021smartblock}.
\vspace{2mm}

Low crop yields and financial difficulties are frequently the result of farmers making poor judgments due to their ignorance of diseases, fertilizers, and new species in agriculture. A service centered on agriculture has been established to help farmers easily and quickly find solutions to their agricultural challenges in order to address this issue. In order to assist farmers in making educated decisions and enhancing crop yields, this service seeks to give information on disease diagnosis, disease prevention, fertilizer recommendations, and novel species  \cite{Plant_diseases}.  Additionally, this website intends to help individuals working in agriculture by offering crucial knowledge on crop disease management \cite{enwiki:1117563996, rahman2024machine}.

\vspace{2mm}
This page offers thorough details on the main ailments affecting different cash crops, vegetable crops, kandal crops, spice crops, fruit crops, flower crops, tree and herbaceous plants, as well as grain crops, pulse crops, oil crops, fiber crops, and cash crops in Bangladesh. It provides thorough explanations of the causes, symptoms, and treatments of each disease, along with pertinent photographs \cite{doi:10.1080/09670878509370984, RahmanWiley}, and covers 534 diseases that can damage a total of 114 crops. If a user has a problem, the expert panel can provide a solution by voice call, video call, or text message. Users can upload photos to receive the best solutions. The user must provide accurate information if he wants to obtain a bank loan. Furthermore, users can pay at checkout if they want to purchase a product. To sum up, the following are the specific contributions of this study:
\vspace{2mm}

\begin{itemize}
    \item Our system \enquote{SMARD} attempts to connect people in need with the availability of public and private agricultural services, easing the procedure and saving time.
    \item We integrate the machine learning classifier and deep learning model into the current system to quickly identify crops diseases.\vspace{1mm}
    
    \item The application is intended to effectively
serve users by giving advice on how to use
products, land, fertilizer, and pesticides to
produce the most crops.\vspace{1mm}

    \item We set-up a text, audio, and video call-based communication mechanism between farmers and an expert panel.\vspace{1mm}
    
    \item Moreover, we can assist farmers in acquiring bank loans, required products, and growing directions for their crops through this system.\vspace{1mm}
    
    \item Finally, our system outperforms the performance of the currently available agricultural-based web applications by achieving 97.3\% classification accuracy and 96\% F1-score in crop disease classification.

\end{itemize}

\section{Literature Review}
We have looked at several websites that offer services relevant to agriculture in order to do a literature review or background investigation. There are surprisingly few businesses in our nation that offer services relating to agriculture \cite{Plant_Disease_Clinic}\cite{Farmers}\cite{AIS}. The agricultural technology currently in use by farmers, such as enhanced seed types, fertilizers, and medical applications, and their effects on crop health and productivity. Furthermore, none of the people who offer this service do it simultaneously with all of the services. An analysis of the present approaches to educating farmers about agricultural technology, taking into account their shortcomings and difficulties. The techniques, outcomes, and constraints of earlier research studies that attempted to employ technology to address agricultural issues are discussed \cite{banglajol, udoy20234sqr, banglasites}.
\vspace{2mm}

A computerized project called Plant Disease Clinic aims to offer efficient fixes for crop disease issues. The most common ailments affecting Bangladeshi-grown grains, pulses, oil seeds, fibers, vegetables, Kendal, spices, fruits, flowers, trees, and herbaceous plants. Causes, symptoms, and treatments are fully described with illustrations \cite{Vegetable_Crops, rahman2023towards}. The U.S. Department of Agriculture (USDA) is in charge of maintaining Farmers, an official website of the federal government of the United States. For farmers, ranchers, and other agricultural producers, the website offers a variety of data, tools, and services. In addition to offering advice on subjects including farm management, conservation, and rural development, it attempts to assist farmers and ranchers in gaining access to the programs, services, and resources provided by USDA agencies. The Bangladesh Ministry of Agriculture is home to the Agricultural Information Service. With the use of the media, the organization's primary goal is to deliver agricultural knowledge and technology to small-scale farmers.\vspace{2mm}

After discovering a number of well-known websites that provide agro-related services, we were able to pinpoint a number of issues with these systems \cite{farmingfuturebd, rahman2023icn}. While some exclusively give agricultural blogs and videos on plant illnesses and their treatments, others include weather updates, the most recent agricultural news, and both. Several websites also provide protection, recovery, and financing services. To provide consumers with even more advantages, we created a system that is more thorough and has extra features.

\section{Methodology}
In this section, we systematically design our framework and propose our solution. 
\subsection{Leaf Object Detection}
The automatic identification and localisation of leaves inside images constitutes the initial stage of our image retrieval methodology. In other words, we train a general leaf detector to produce bounding box coordinates and confidence ratings for every leaf region contained within them given an input leaf image. The phenotypic traits of many plant kinds clearly differ from one another, with each leaf exhibiting a unique texture and shape. For many detection systems, finding leaves in a complicated environment with accuracy and timeliness is a major difficulty. For this we use YOLOv5 \cite{Jocher2020} which is is known deep learning object detection model.\vspace{2mm}

\subsection{Dataset Collection}
\begin{figure}[h!]
  \centering
    \includegraphics[width=0.55\textwidth]{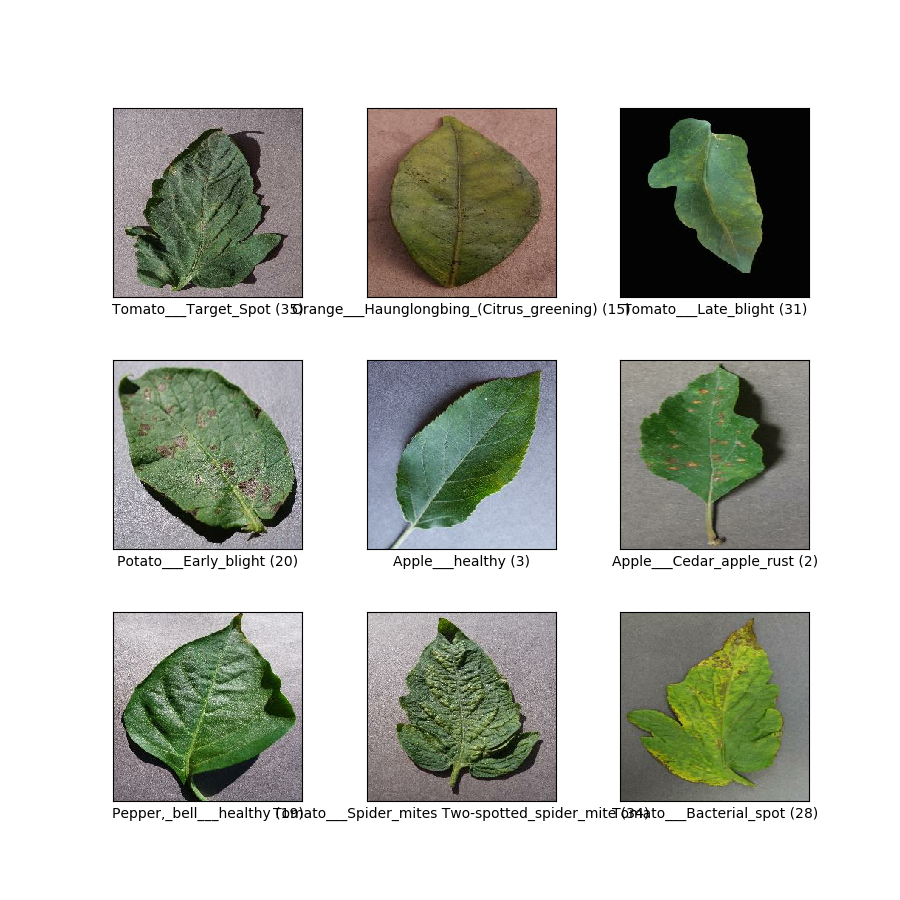}
    \caption{PlantVillage dataset}
	\label{plant_disease}
\end{figure}

The PlantVillage dataset is currently regarded as one of the largest publicly accessible collections of well-curated plant leaf images for the detection of diseases. It contains 54,309 images of healthy and diseased leaves from fourteen harvests that have been categorized by experts in plant pathology. A total of 16,011 pictures were used in the studies, which included 9 different leaf sickness classes and 1 normal class. This dataset includes samples of leaves with differing degrees of disease infection \cite{EasyChair:5023}. A picture from different classes labeled different diseases in the database is displayed in Figure. \ref{plant_disease}.

\vspace{2mm}
\subsection{Feature Extraction and Classification}
\begin{figure*}[h!]
  \centering
    \includegraphics[width=0.8\textwidth]{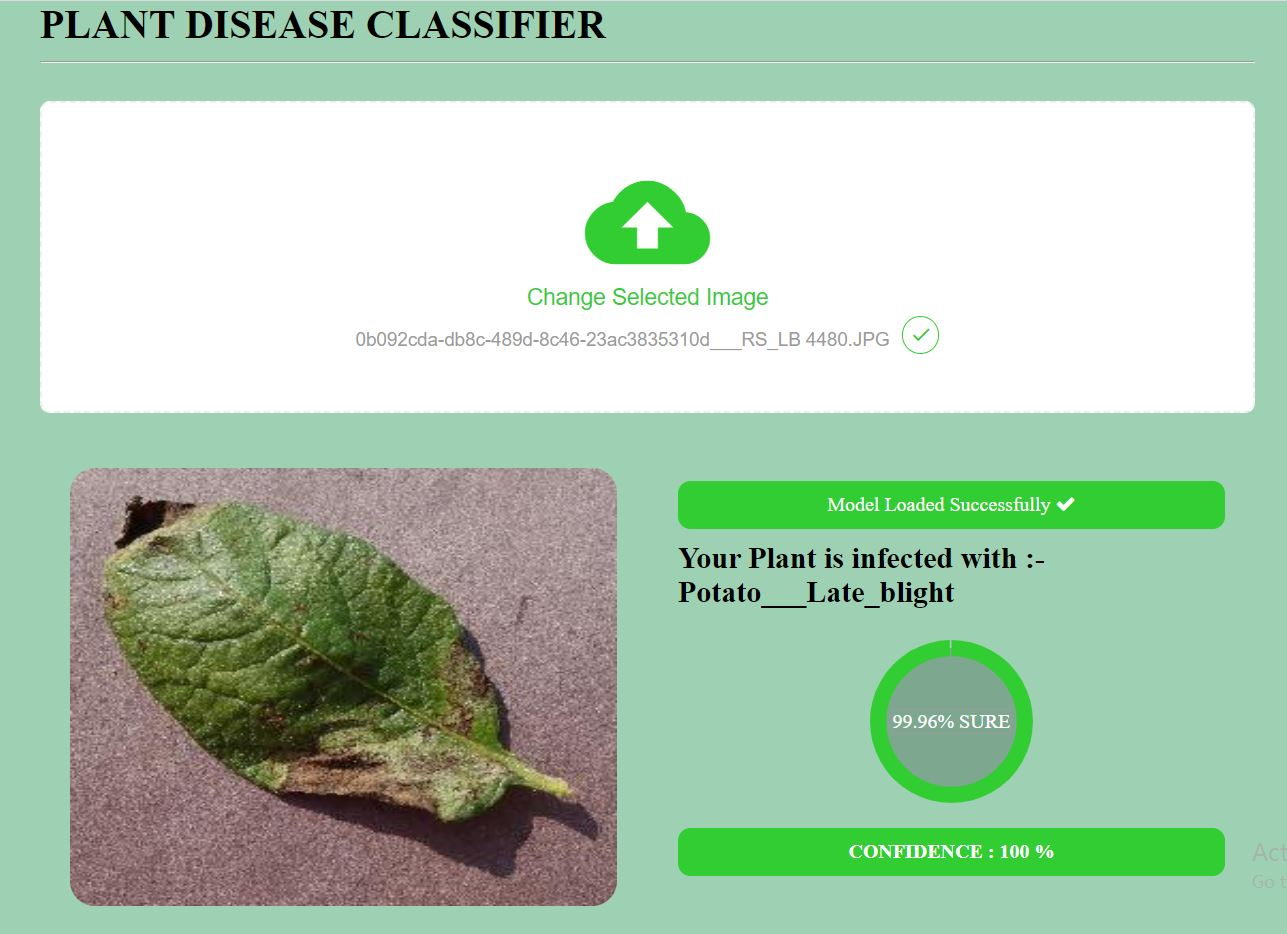}
    \caption{Implementation of Plant disease classification using pre-trained CNN Network }
	\label{fig: name}
\end{figure*}

Our feature extraction network attempts to get robust and discriminating semantic data from the leaf picture, which is then transferred to a fixed-dimension feature vector. To achieve results for leaf disease recognition, this feature vector will finally be matched with the same claimed identity vector. It is clear that the accuracy of system retrieval is largely determined by the quality of extracted characteristics. In order to increase the model's stability and precision, we suggested a feature extraction network that combines metric learning and conventional supervised classification prediction. We used the best reflection of pre-trained CNN models in our \enquote{SMARD} project \cite{jung2023construction, rahman2023federated}. Figure. \ref{fig: name} shows the plant disease classification process of our \enquote{SMARD} project.

\begin{figure*}[h!]
  \centering
    \includegraphics[width=.99\textwidth]{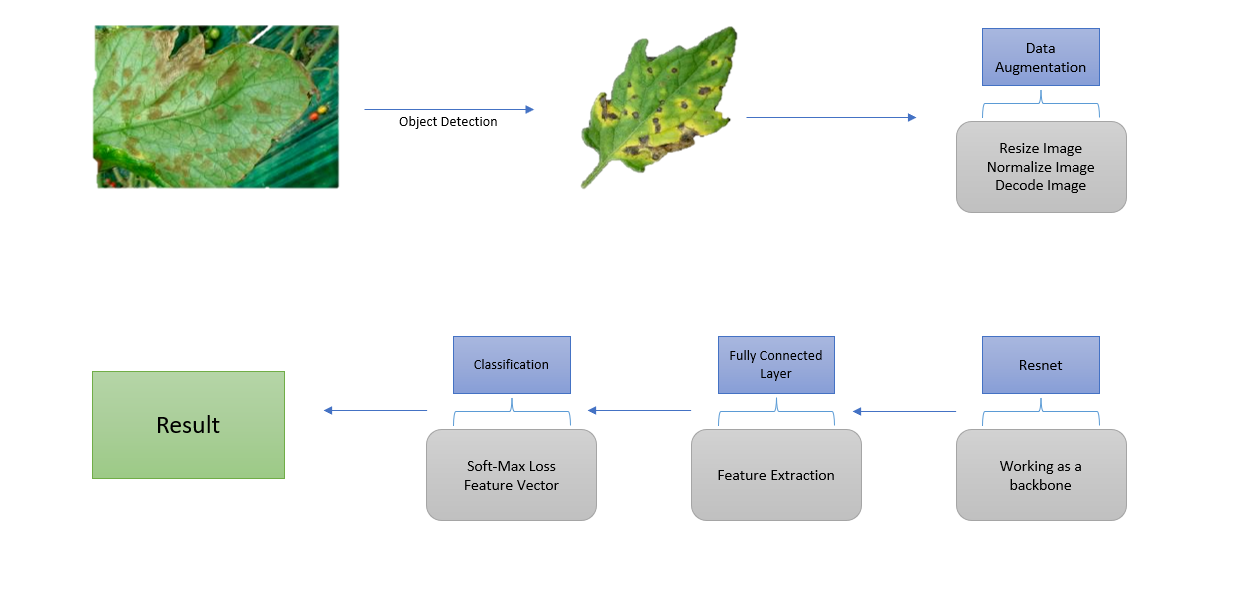}
    \caption{Crops Disease Detection Pipeline of \enquote{SMARD}}
	\label{Network_Model}
\end{figure*}

\begin{figure*}[h!]
  \centering
    \includegraphics[width=0.96\textwidth]{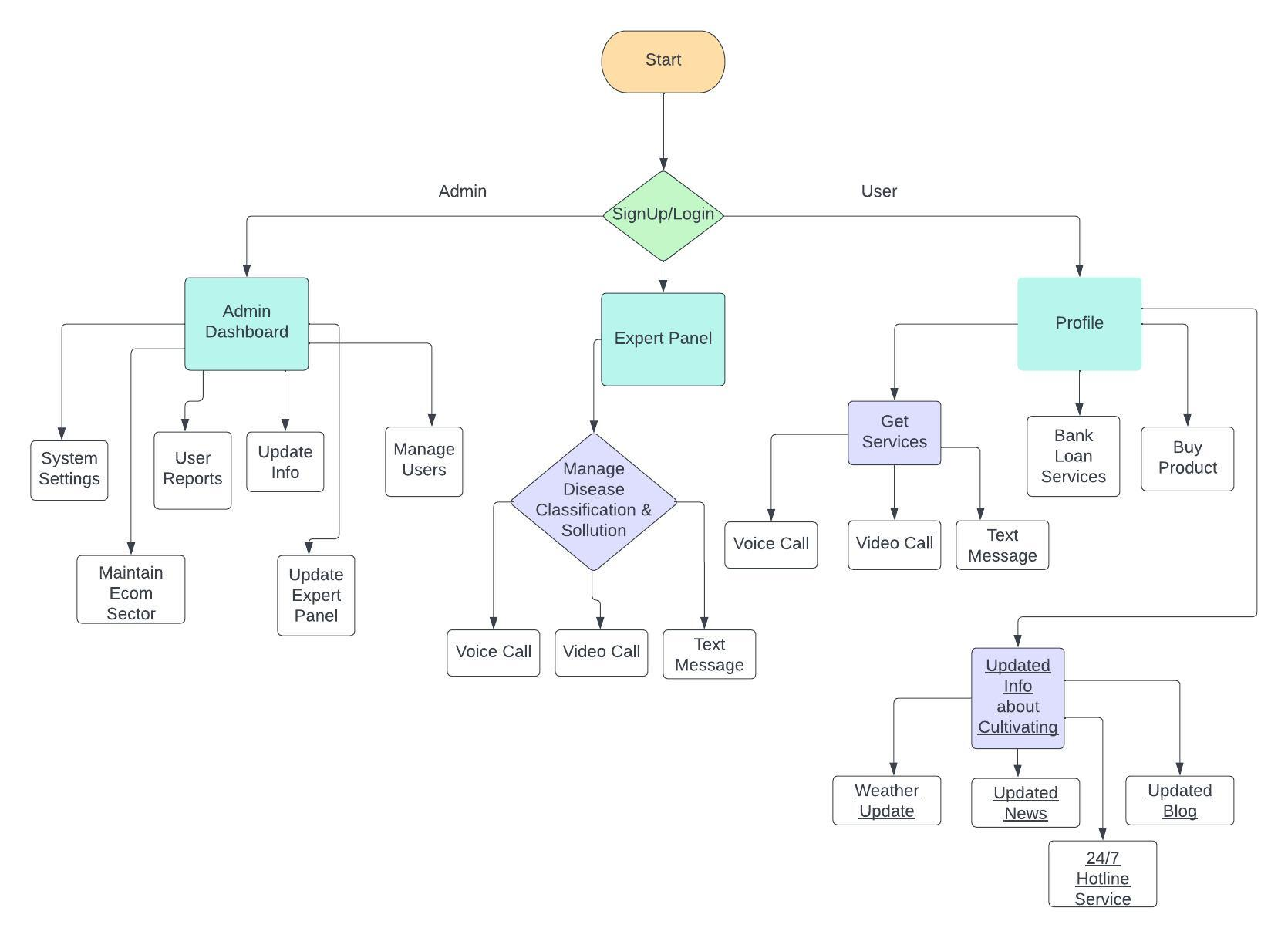}
    \caption{Proposed workflow of our web application based system \enquote{SMARD}}
	\label{fig: 1}
\end{figure*}

\subsection{Network Architecture}
We employ typical data augmentation techniques used in classification problems, such as image decoding, normalizing image pixel values, randomly flipping images horizontally, and cropping photos to 224 x 224 in size. We used CNN as our backbone in the framework. Deep features typically retrieved from most backbone networks have ultra-high dimensions, which means that using them directly will reduce the effectiveness of vector search in picture retrieval and need more calculations. Therefore, the fully connected layer is implemented in the feature extraction module to condense the output of the backbone network into prefixed-size feature maps for our jobs. to train with the images
To classify images of leaf disease, we have used a fully-connected model. The feature vector size serves as its input dimension, and the training set's sample category numbers serve as its output dimension.
Figure. \ref{Network_Model} shows the structure of our feature extraction network and also reflects the pipeline of crops disease detection process.

\subsection{Proposed System}
After reviewing various works related to agricultural services, we have found limitations. Some offer weather updates and agricultural news, while others only offer blogs and videos on plant diseases. A few provide loan services, protection, and recovery solutions. We aim to develop a more comprehensive system with additional features for greater benefit to the users. So we try to provide a complete package for the plant disease classification. A flowchart has been provided in the Figure. \ref{fig: 1} to provide an overview of our project.

\section{Performance Analysis and Discussion}
One set of evaluation criteria used to assess the effectiveness of the proposed pipeline is the Mathew correlation coefficient (MCC), which includes precision, specificity, F1-score, accuracy, sensitivity, and precision. We present and discuss the experimental results for the classification outcome and see that we score almost 97.3\% accuracy and 96\% F-1 score.

\section{Limitations and Future Scope}
It is impractical to assert that a project is fully free of any constraints because all initiatives have intrinsic limitations. We made every effort to include all important functionality in our project, however, we do acknowledge that there are certain restrictions. It may be challenging for people to use the system properly without having a fundamental understanding of technology and computer processes. To get the maximum performance out of this project, more work is needed to completely integrate the image processing functionality.\vspace{2mm}

In order to effectively control various fish diseases, identify acceptable ponds, and find adequate food sources for the fish, it is important to forge connections within the fisheries sector. We will meet the demands of cattle in a variety of ways, such as ensuring their well-being, choosing suitable housing, and providing them with wholesome food. Additionally, we will gradually increase the security of our system and work to make it more flexible and user-friendly. Additionally, we intend to create a mobile application for both the iOS and Android operating systems.

\section{Conclusion}
Bangladesh is primarily dependent on agriculture, which is experiencing a number of difficulties as a result of population increase, a lack of available farmland, ecological imbalances, and crop pests. To prevent crop damage, farmers use a lot of pesticides, which raises production costs and has a bad effect on the environment. As a result, the agriculture sector is rapidly deteriorating. To swiftly diagnose crop illnesses, we incorporate the machine learning classifier and deep learning model into the current system. By attaining 97.3\% classification accuracy and 96\% F1-score in crop disease classification, our system outperforms the performance of the currently available agricultural-based web applications. \vspace{2mm}

Additionally, our project \enquote{SMARD} provides a range of services to assist farmers in effectively resolving their agricultural problems in order to address these concerns. These services include offering thorough information on issues like crop and seed diseases, the best way to utilize fertilizer and pesticides, choosing the right area and soil, and high-yield farming. Higher yields and earnings will result from farmers having access to agricultural products and the option of obtaining bank loans. They can also learn about new seed kinds, illnesses, and the proper handling and timing of fertilizer applications. With the help of our efforts, Bangladeshi farmers would be able to significantly boost the country's economy.

\bibliographystyle{IEEEtran}
\bibliography{Simple}

\end{document}